\documentclass[useAMS,usenatbib]{mn2e} 
\usepackage{epsfig,lscape,amsmath,amssymb} 
\usepackage{graphicx,longtable, times, color}

\DeclareGraphicsExtensions{.ps,.eps,.eps.gz,.epsi}

\newcommand\aj{AJ}  %
\newcommand\aap{A\&A}  %

\begin{document}

\title[]{Relative Flux Calibration of the LAMOST Spectroscopic Survey of the Galactic Anti-center}

\author[Xiang et al.]{M. S. Xiang$^{1}$\thanks{E-mail: xiangmaosheng@pku.edu.cn}, 
        X. W. Liu$^{1,2}$\thanks{E-mail: x.liu@pku.edu.cn}, H. B. Yuan$^{2}$, 
        Z. Y. Huo$^{3}$, Y. Huang$^{1}$, Y. Zheng$^{1}$, H. W. Zhang$^{1}$, 
        \newauthor B. Q. Chen$^{1}$, H. H. Zhang$^{1}$, N. C. Sun$^{1}$, C. Wang$^{1}$, Y. H. Zhao$^{3}$, 
        J. R. Shi$^{3}$, A. L. Luo$^{3}$, 
        \newauthor G. P. Li$^{4}$, Z. R. Bai$^{3}$, Y. Zhang$^{4}$, Y. H. Hou$^{4}$, H. L. Yuan$^{3}$, G. W. Li$^{3}$
\\ \\ $1$ Department of Astronomy, Peking University, Beijing 100871, P. R. China \\
$2$ Kavli Institute for Astronomy and Astrophysics, Peking University, Beijing 100871, P. R. China \\
$3$ Key Laboratory of Optical Astronomy, National Astronomical Observatories, Chinese Academy of Sciences, Beijing 100012, P. R. China \\
$4$ Nanjing Institute of Astronomical Optics \& Technology, National Astronomical Observatories, Chinese Academy of Sciences, Nanjing 210042, P. R. China \\ }
\date{Received:}

\maketitle

\label{firstpage}

\begin{abstract}{
         We have developed and implemented an iterative algorithm of flux calibration 
         for the LAMOST Spectroscopic Survey of the Galactic anti-center (LSS-GAC). 
         For a given LSS-GAC plate, the spectra are first processed with a set of 
         nominal spectral response curves (SRCs) and used to derive initial stellar 
         atmospheric parameters (effective temperature $T_{\rm eff}$, surface gravity 
         log\,$g$ and metallicity [Fe/H]) as well as dust reddening $E(B-V)$ of all 
         targeted stars. For each of the sixteen spectrographs, several F-type stars 
         of good signal-to-noise ratios (SNRs) are then selected as flux standard stars 
         for further, iterative spectral flux calibration. Comparison of spectrophotometric 
         colours, deduced from the flux-calibrated spectra, with the photometric 
         measurements yields average differences of 0.02$\pm$0.07 and $-$0.04$\pm$0.09\,mag 
         for the $(g-r)$ and $(g-i)$, respectively. The relatively large negative 
         offset in $(g-i)$ is due to the fact that we have opted not to correct 
         for the telluric bands, most notably the atmospheric A-band in the wavelength 
         range of $i$-band. Comparison of LSS-GAC multi-epoch observations of duplicate 
         targets indicates that the algorithm has achieved an accuracy of about 10 per cent 
         in relative flux calibration for the wavelength range 4000 -- 9000\,\AA.
         The shapes of SRC deduced for the individual LAMOST spectrographs are found 
         to vary by up to 30 per cent for a given night, and larger for different nights, 
         indicating that the derivation of SRCs for the individual plates is essential 
         in order to achieve accurate flux calibration for the LAMOST spectra.}

\end{abstract}

\begin{keywords} Galaxy: disk -- Galaxy: general -- techniques: spectroscopic -- survey  
\end{keywords}

\section{Introduction}          
\label{sect:intro}

The LAMOST Spectroscopic Survey of the Galactic anti-center
(LSS-GAC; Liu et al. 2014) is a major component of
the on-going Galactic spectroscopic surveys with the Large
Sky Area Multi-Object Fiber Spectroscopic Telescope (LAMOST,
also named the $Guo\,Shoujing$ telescope; Cui et al. 2012).
The LSS-GAC will collect millions of spectra of Galactic stars of all
colours in a contiguous sky area of about 3,400 square
degrees centered on the Galactic anti-center, from Galactic
longitude 150$^{\rm o}$ to 210$^{\rm o}$ and latitude from $-30^{\rm o}$
to $+30^{\rm o}$., and yield a statistically complete sample of
stars with accurate stellar parameters, including radial velocity  ($V_{\rm r}$), 
effective temperature ($T_{\rm eff}$), surface gravity (log\,$g$) 
and metallicity ([Fe/H]), to an unprecedented sampling depth and 
density for the purpose of revealing the formation and evolution
history of the Milky Way. Preceded by one-year-long Pilot Surveys,
the LAMOST Regular Surveys were initiated in October 2012.

Obtaining accurate calibration of the stellar spectral energy 
distribution (SED) is a prerequisite for robust stellar spectral 
classification and reliable stellar (atmospheric) parameter 
determinations, and is thus an essential step to 
the fulfillment of the scientific goals of LSS-GAC. 
The recorded spectral signal as a function of wavelength of a star 
is affected by a number of processes that occur between the 
star and observer, including the interstellar and Earth atmospheric 
extinction and the spectral throughput of the telescope and 
instruments. The purpose of the so-called spectral flux calibration 
is to obtain accurate stellar SEDs above the Earth atmosphere, 
and this involves the characterization and removal of the 
atmospheric and instrumental effects as a function of wavelength. 
The effects of extinction by the interstellar dust grains are 
typically left to a later stage of the analysis, as they vary 
from star to star (Chen et al. 2014; Yuan et al. 2014a).

In astronomical spectroscopy, spectral flux calibration of target 
objects is generally achieved by obtaining separate measurements 
of spectrophotometric standard stars (e.g. Oke 1990; Hamuy et al. 1992, 1994) 
on the same night with the same instrumental setup. For modern large 
scale multiplex spectroscopic surveys employing multiple spectrographs 
and hundreds of fibers, obtaining separate measurements of standard 
stars for each night and each spectrograph/fiber becomes extremely 
costly in term of observing time and essentially impossible, and 
a different strategy has to be adopted. In the case of 
the Sloan Digital Sky Survey (SDSS; York et al. 2000), which 
primarily targets extragalactic objects at high 
Galactic latitudes that suffer from relatively small interstellar 
extinction, F turn-off stars within the field of view (FoV), 
pre-selected based on the photometric colours and observed 
simultaneously with the targets are used to calibrate the spectra 
(Stoughton et al. 2002; Yanny et al. 2009), 
assuming that the intrinsic SEDs of F turn-off stars are well 
determined by the theoretical stellar atmospheric models and the 
effects of the interstellar extinction can be characterized and 
removed using the all-sky extinction map of 
Schlegel, Finkbeiner \& Davis (1998; SFD98 hereafter). 
In the LAMOST case, Song et al. (2012) have proposed a method 
to calibrate the spectra. However, their method does not 
incorporate dereddening of the flux calibration standard stars, 
thus can not be applied to high extinction regions. 
For the LSS-GAC, since the majority of targets are disk stars 
at low Galactic latitudes in the direction of Galactic anti-center (GAC), 
the targets are subject to significant, unknown amounts of 
the interstellar extinction. As such, it is difficult 
to pre-select flux calibration standard stars based on the photometric 
colours alone. Ideally, one can use the spectral response curves 
(SRCs) deduced from high Galactic latitude fields observed on 
the same night. However, as we shall show later in the current work, 
the SRCs of the individual LAMOST spectrographs are found to vary 
significantly not only from night to night, but also in a given night. 
 
Fortunately, one primary goal of the LSS-GAC is to 
sample stars of all colours (spectral types). For this purpose, 
a simple yet non-trivial target selection algorithm is developed 
where sample stars are uniformly selected spatially as well as 
on the ($r$, $g-r$) and ($r$, $r-i$) Hess diagrams 
(Liu et al. 2014; Yuan et al. 2014b). 
Thus for each FoV observed, there are always some stars targeted 
that are suitable for the purpose of spectral flux calibration, 
including F turn-off stars, in spite the fact that they 
cannot be easily identified photometrically 
due to the effects of interstellar reddening. 

In this work, we have developed and implemented an iterative
algorithm to derive SRCs and flux-calibrate the LSS-GAC spectra.
We start with wavelength-calibrated, flat-fielded and
sky-subtracted 1-D spectra processed with the LAMOST 2-D
pipeline (Luo et al. 2012; Bai et al., in prep), and aim to
obtain accurately flux-calibrated spectra recorded in both
the blue and red arms of spectrographs.
We concentrate on the relative rather than absolute flux calibration. 
Given good relative spectral flux calibration, absolute calibration 
can be achieved by scaling the spectra to broad-band photometric measurements. 
We estimate the accuracy of our method by comparing results 
of multi-epoch observations of duplicate targets, as well as
by comparing stellar colours yielded by the calibrated spectra 
with those from the photometric measurements. 
SRCs deduced for individual plates observed in different
nights are compared to investigate the possible variations
of SRCs with time, useful for the future improvement of the 
strategy of flux calibration for LAMOST spectra.
LSS-GAC spectra calibrated with our method are used by the LAMOST Stellar
Parameter Pipeline at Peking University (LSP3; Xiang et al. 2014 submitted) 
for stellar parameter determinations. The spectra have also been used
to search for white dwarfs (WDs) and white dwarf – main sequence (WDMS)
binaries (Rebassa-Mansergas et al. 2014 submitted; Ren et al. 2014).
Following the data policy of LAMOST surveys, the calibrated 
spectra in fits format will be publically available as value-added products 
along with the first data release of LAMOST (LAMOST DR1; Bai et al., in prep), 
currently scheduled in December, 2014, and can be accessed 
via http://162.105.156.249/site/LSS-GAC-dr1/, along with a description file. 

The paper is organized as follows. In Section\,2, we present our 
flux-calibration algorithm. The results and discussions are presented 
in Section\,3. Section\,4 is the summary. 

\section{Methodology}
LAMOST\footnote{http://www.lamost.org/website/en/}, also known as 
Wang-Su Reflecting Schmidt Telescope or Guoshoujing Telescope, 
is a new type of wide field telescope with a large aperture. 
It has an effective aperture of $\sim$\,4\,m and a (circular) FoV of 
5$^{\rm o}$ in diameter (Wang et al. 1996; Cui et al. 2012). 
At the focal plane, 4000 robotic optical fibers of an entrance aperture
of 3.3\,arcsec projected on the sky relay the light to 16 low-resolution
spectrographs, 250 fibers each. To increase the spectral resolution, 
slit marks of width 2/3 the fiber diameter are used, yielding a 
resolving power of $R$\,$\sim$\,1,800 (Deng et al. 2012; Liu et al. 2014).
Each spectrograph has two arms to collect the collimated light beam. 
The wavelength coverage is 3700 -- 5900\,{\AA} in the blue arm 
and 5700 -- 9000\,\AA\ in the red arm. The final spectra  
spectra cover the wavelength range from 3700 to 9000\,{\AA}. 
In each arm of a given spectrograph, a 4K$\times$4K CCD, with a 
squared pixel size of 12\,$\mu$m, is used to record light 
signal (Cui et al. 2012). One CCD pixel corresponds to about 0.56\,{\AA} 
in the blue, and about 0.82\,{\AA} in the red.

Raw data of the LSS-GAC are first processed with the LAMOST 2-D 
pipeline (Bai et al., in prep), which includes the steps of fiber tracing, 
wavelength calibration, flat fielding, sky subtraction, to produce 
wavelength-calibrated, flat-fielded and sky-subtracted 1-D spectra.
The spectra are then flux-calibrated via the process introduced in 
the current work. Note that in both the default LAMOST 2-D pipeline 
and in the flux calibration algorithm presented here, the data are 
reduced spectrograph by spectrograph.

\subsection{Overview}
The flux calibration of LSS-GAC spectra is implemented with an
iterative process. As mentioned above, we start with the
wavelength-calibrated, flat-fielded and sky-subtracted 1-D spectra.
A flowchart of the flux calibration is illustrated in Fig.\,1. 
For a given plate, which includes data collected from 32 CCDs 
of the 16 spectrographs, the uncalibrated 1-D spectra are first processed 
with a set of nominal SRCs, deduced from observations of high Galactic 
latitude fields adopting F turn-off stars as standards, following 
the practice of the SDSS. 
The resultant spectra are then used to derive first estimates of the stellar
parameters ($V_{\rm r}$, $T_{\rm eff}$, log\,$g$ and [Fe/H]) for all stars
targeted by the plate using the LAMOST Stellar Parameter Pipeline
at Peking University (LSP3; Xiang et al. 2014, XLYH14 hereafter). 
Based on those deduced stellar parameters, for each of the 16 spectrographs,
several F-type stars of good spectral signal-to-noise rations (SNRs) and 
well determined parameters are selected as flux calibration standards, 
which are then used to derive an updated set of SRCs by comparing their
observed spectra with synthetic ones interpolated to the
corresponding parameters using the spectral library of Munari et al.
(2005). In deriving the SRCs, piecewise low-order polynomials
are used to fit the ratios of the observed and synthetic spectra,
with the latter reddened with a colour excess $E(B-V)$
estimated by comparing the measured and synthetic photometric
colours, assuming the Fitzpatrick (1999) extinction law for a 
total-to-selective extinction ratio $R_V = 3.1$.
Once calibrated, the blue- and red-arm spectra are pieced together,
and spectra from consecutive individual exposures are co-added.
Spectra thus reprocessed using the newly derived SRCs are then
used to update the stellar parameter estimates, and the above process
is repeated. Typically, the results 
(e.g. as represented by the stellar parameters yielded by the
processed spectra) converge after 2 -- 3 iterations. 
Note that since each spectrograph covers a FoV of approximately 1\,deg. only,
we have ignored the small differences of atmospheric extinction
for individual stars within the FoV of the spectrograph. The errors 
thus introduced to the SEDs are less than $\sim$\,2 per cent for the whole 
spectral wavelength range even for a zenith distance of 60\,deg. (airmass $\sim$\,2.0).

\begin{figure}
\centering
\includegraphics[width=8cm]{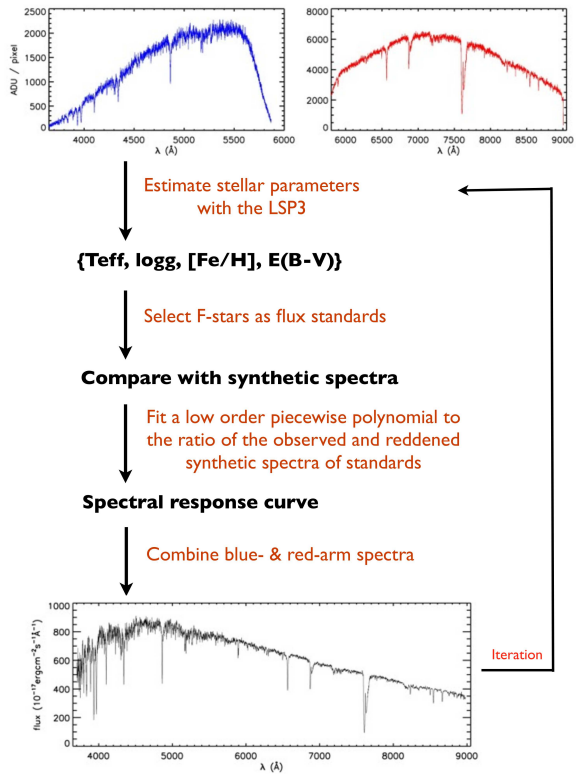}
\caption{Flowchart of the flux calibration.}
\label{Fig1}
\end{figure}

\subsection{Stellar atmospheric parameters}
As outlined above, we calculate the intrinsic SEDs of standard stars 
by interpolating the spectral library of Munari et al. (2005) to the 
desired stellar atmospheric parameters, i.e, $T_{\rm eff}$, log\,$g$ and [Fe/H],
of the stars. The accuracy of the stellar atmospheric
parameters, in particular $T_{\rm eff}$, which determines the
shape of the stellar SED, thus has a direct impact on the 
accuracy of the deduced SRCs. We determine the stellar atmospheric
parameters from the spectra using the LSP3, which derives
the parameters by matching the target spectra
with the MILES library (S\'anchez-Bl\'azquez et al. 2006).
As discussed in detail in XLYH14, for FGK stars, an accuracy of
150\,K, 0.25\,dex and 0.15\,dex has been achieved by the LSP3 for
$T_{\rm eff}$, log\,$g$ and [Fe/H], respectively, given a spectral 
SNR higher than 10. Note that as in XLYH14, 
unless specified otherwise the SNR of spectrum refers to 
the median value per pixel for the wavelength range of 4600 -- 4700\,{\AA}. 
For LAMOST spectra, one pixel at 4650\,{\AA} corresponds to 1.07\,{\AA}. 
For $T_{\rm eff}$ between 5,750 and 6,750\,K, the temperature
range occupied by most F stars, an uncertainty of 150\,K in
$T_{\rm eff}$ leads a maximum error in the shape of SED
of the star of 9 and 3 per cent for the wavelength ranges covered by 
the blue and red-arm spectra, respectively, and a maximum error 
of 12 per cent for the whole wavelength range.
The errors introduced by the uncertainties in log\,$g$ and [Fe/H] 
are marginal, on the level of 1 per cent, and thus can generally be ignored. 

\subsection{Standard stars}

For each of the 16 spectrographs, we select stars with
5750\,$\le T_{\rm eff}\le$\,6750\,K, log\,$g$\,$\ge$\,3.5\,(cm\,s$^{-2}$) 
and $-1.0\leq {\rm [Fe/H]}\leq$ 0.5\,dex as flux calibration standards. 
The atmospheric parameters of such stars are well determined by the LSP3
(XLYH14) and their intrinsic SEDs are well modeled by the stellar 
atmospheric models. Benefited from the large fiber number (4000) 
of LAMOST and the target selection algorithm employed by the LSS-GAC
(randomly and uniformly on the colour-magnitude Hess diagrams; Liu et al. 2014;
Yuan et al. 2014b submitted), almost in all cases there are sufficient 
stars within the above parameter space that are targeted in a given 
spectrograph and thus can be used as flux-calibration standards.
Fig.\,2 shows the distribution of the number of
standards per spectrograph, as well as the distribution of
$T_{\rm eff}$ of the selected standard stars, for 602 spectral
plates collected by June 2013. 
Most of the plates are from the LSS-GAC except for a few 
from the Galactic spheroid survey (Deng et al. 2012) 
for the purpose of comparing the SRCs of plates of high and low 
Galactic latitudes. The Figure shows that 
71 per cent of the spectrographs have more than 4 standard stars, 
and only $\sim$\,13 per cent of them have less than 3 standards.
Note we have set an upper limit of 10 standards per spectrograph.
The number of usable standard stars are in general limited by the SNRs. 
Plates with few usable standards belong to either the Medium or 
the Faint plates (Liu et al. 2014; Yuan et al. 2014b submitted) observed under 
unfavorable weather conditions such that the plates have not 
reached the desired SNRs. We have set a floating SNR cut when 
selecting standards in order to achieve a balanced number of stars. 
More than 90 per cent of the selected standards have SNR higher than 15 
for a single exposure (most observations consist of two to three consecutive 
exposures of equal integration time).
There are 67 plates for which the SNRs are too low to 
select standard star. Those plates are processed using a set of nominal SRCs. 
  
\begin{figure}
\centering
\includegraphics[width=8cm]{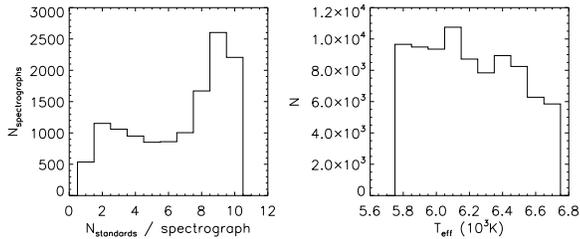}
\caption{Histograms of the numbers of standard stars selected 
         per spectrograph (left) and the effective 
         temperatures of the standard stars (right).}
\label{Fig2}
\end{figure}

\subsection{Spectral response curves}

Let $F_{\rm obs}(\lambda)$ and $F_{\rm int}(\lambda)$ denote 
the measured and intrinsic spectral flux density, we have,
\begin{equation}
F_{\rm obs}(\lambda) = 10^{-0.4[A_{\rm i}(\lambda) +
A_{\rm a}(\lambda)X]}R(\lambda)F_{\rm int}(\lambda)
\end{equation}
where $A_{\rm i}(\lambda)$ is the interstellar extinction 
by dust grains, $A_{\rm a}(\lambda)$ the Earth atmospheric 
extinction per unit airmass, $X$ the airmass and $R(\lambda)$ 
the telescope and instrumental SRC. $F_{\rm int}(\lambda)$, 
in units of ergs\,cm$^{-2}$\,s$^{-1}$\,{\AA}$^{-1}$, is calculated 
by interpolating the spectral library of Munari et al. (2005) 
to the desired $T_{\rm eff}$, log\,$g$, 
and [Fe/H] of the standard star in concern, normalized to 
the $r$-band photometric magnitude measured by 
the Xuyi Schmidt Telescope Photometric Survey of the Galactic 
Anti-center (XSTPS-GAC; Liu et al. 2014). For Very Bright (VB) 
plates that target stars of $r \leq 14$\,mag. 
that are saturated in the XSTPS-GAC survey, the 2MASS (Skrutskie et al. 2006) 
$J$-band magnitude is used to normalize $F_{\rm int}(\lambda)$ instead. 

The synthetic spectra of 1\,{\AA} dispersion from the library of Munari et al. (2005) 
are used. The spectra are degraded to the LAMOST spectral resolution 
by convolving with a Gaussian function. 
Although spectra of various micro-turbulent and rotation 
velocities are provided by the library, only those of a constant 
micro-turbulent velocity of 2.0\,km\,s$^{-1}$ and a zero rotation velocity 
are used, as these two parameters have little effects on the  
SED at a given temperature. In addition, only spectra calculated 
using the new Opacity 
Distribution Function (ODF), flaged as `NW' in the library, are used. 
If ${\rm [Fe/H]} < -1.0$\,dex, then spectra of ${\rm [}\alpha/{\rm Fe]} = 0.4$\,dex 
are adopted, otherwise those of ${\rm [}\alpha/{\rm Fe]} = 0.0$\,dex are adopted. 
The spectra are linearly interpolated to the desired 
$T_{\rm eff}$, log\,$g$ and [Fe/H]. 

The interstellar extinction $A_{\rm i}(\lambda)$ can be expressed as, 
\begin{equation}
A(\lambda) = [R_V + k(\lambda-V)] \times E(B-V)
\end{equation}
where $R_V \equiv A(V)/E(B-V)$ is the ratio of total to 
selective extinction, $E(B-V) = A(B) - A(V)$ the colour excess of $B$ and $V$ bands, 
and $k(\lambda-V)$ = $E(\lambda-V)/E(B-V)$ the extinction curve. 
We adopt the extinction curve of Fitzpatrick (1999), assuming $R_V=3.1$. 
The value of $E(B-V)$ of a selected flux calibration standard star is determined 
by comparing the photometric colours of XSTPS-GAC $g$, $r$ and $i$ bands in the 
optical and the 2MASS $J$, $H$ and $K_{\rm s}$ bands in the near infrared of 
the standard star with those predicted by the synthetic spectrum of the same stellar 
atmospheric parameters ($T_{\rm eff}$, log\,$g$, [Fe/H]) as the standard star of concern, 
with the latter interpolated from the library of Castelli \& Kurucz (2004). 
The derived $E(B-V)$ have typical uncertainties of about 0.04\,mag (Yuan et al. 2014b submitted). 
For stars that are saturated in the XSTPS-GAC survey, only the 2MASS colours are 
used, with the resultant $E(B-V)$ suffering from slightly larger uncertainties 
($\sim$\,0.06\,mag.). 

As for the atmospheric extinction $A_{\rm a}(\lambda)$, we adopt the 
extinction coefficients retrieved from the website\footnote{http://www.xinglong-naoc.org/gcyq/}  
deduced from the multi-medium-band photometry of the Beijing-Arizona-Taiwan-Connecticut (BATC) 
survey (cf. Fan et al. 1996). 
Note that the potential uncertainties and temporal variations of the 
extinction coefficients generally do not have an impact on the final 
accuracy of flux-calibration, though they do affect the shapes of 
SRCs deduced and their possible temporal variations inferred consequently (cf. \S\,{3.3}). 

\begin{figure*}
\centering
\includegraphics[width=18cm]{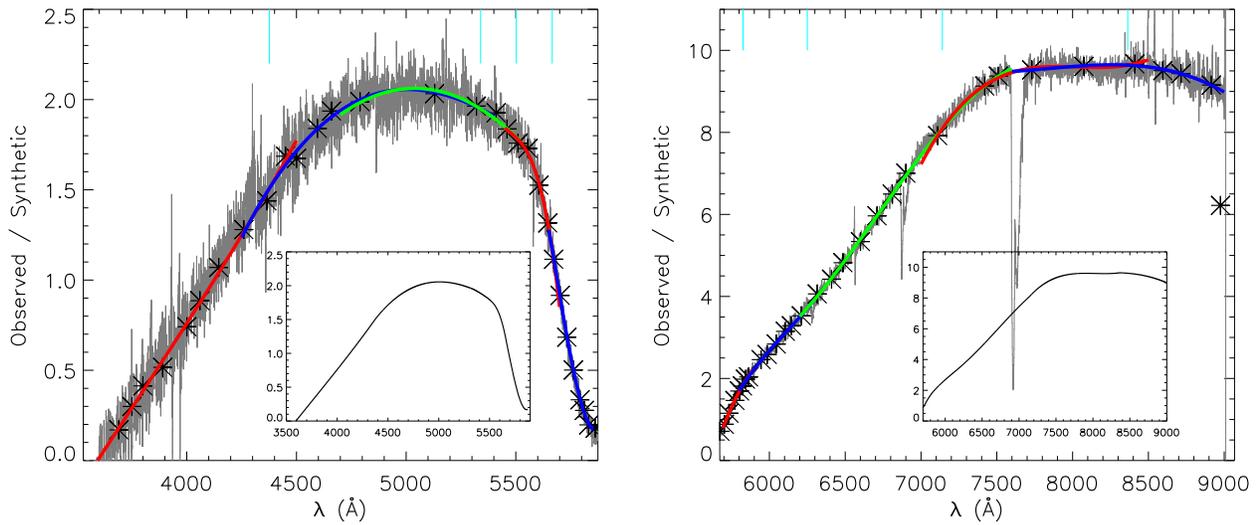}
\caption{Examples of SRC fitting for the blue- (left) and red-arm (right) 
         spectra. The grey lines are ratios of the observed flux density 
         (in units of ADU\,cm$^{-2}$\,s$^{-2}$\,{\AA}$^{-1}$) after corrected 
         for the Earth atmospheric extinction and the flux density of the 
         synthetic spectrum (in units of erg\,cm$^{-2}$\,s$^{-1}$\,\AA$^{-1}$) 
         after dereddened for the interstellar extinction.        
         The ratios have been scaled by a factor of $10^{-17}$. 
         Asterisks in red are median values of spectral regions 
         clear of prominent spectral features or artifacts that are 
         used to generate the SRCs (see text for detail). 
         The thick lines in colours represent piecewise polynomial fits 
         to the asterisks. The vertical lines in cyan indicate 
         positions where adjacent fits to individual wavelength regimes are 
         pieced together. 
         The black curves in the inserts represent the final SRCs.}
\label{Fig3}
\end{figure*}
SRCs are assumed to be smooth functions of wavelength. To derive SRCs, 
we apply a low-order piecewise polynomial fitting to the ratios 
of the observed and intrinsic (synthetic) spectra of the standard 
stars after correcting the latter for the effects of the interstellar and 
Earth atmospheric extinction. Fig.\,3 shows examples of SRC fitting. 
We define a series of clean spectral regions avoiding the prominent 
stellar absorption features and telluric absorption bands. 
We opt not to remove the latter from the flux-calibrated LSS-GAC spectra 
in the current implementation of the pipeline version (cf. \S\,{2.5}). 
The median values of data in those clean regions, indicated by asterisks 
in Fig.\,3, are used for the SRC fitting. For each standard star, both 
the blue- and red-arm spectra are divided into 5 wavelength regimes, 
and each regime is fitted with a 2nd or 3rd-order polynomial, 
as represented by the thick lines in colours in the Figure. 
The vertical solid lines in cyan indicate where the fits of adjacent 
regimes are jointed together to produce the final SRC for the whole wavelength range. 
The wavelengths of the joint points are not fixed 
but determined by the crossing wavelengths of fits of the adjacent 
spectral regimes. We assume that the differences in sensitivity of the individual 
fibers have been well corrected via flat-fielding and thus the 250 fibers 
of a given spectrograph share a single SRC. The final SRC is 
adopted as the SNR-weighted average of SRCs deduced from the individual 
selected standard stars observed with the spectrograph of concern. 
Fig.\,4 shows the SRCs of the 16 LAMOST 
spectrographs derived from one plate observed on October 27, 2011. 
\begin{figure*}
\centering
\includegraphics[width=18cm]{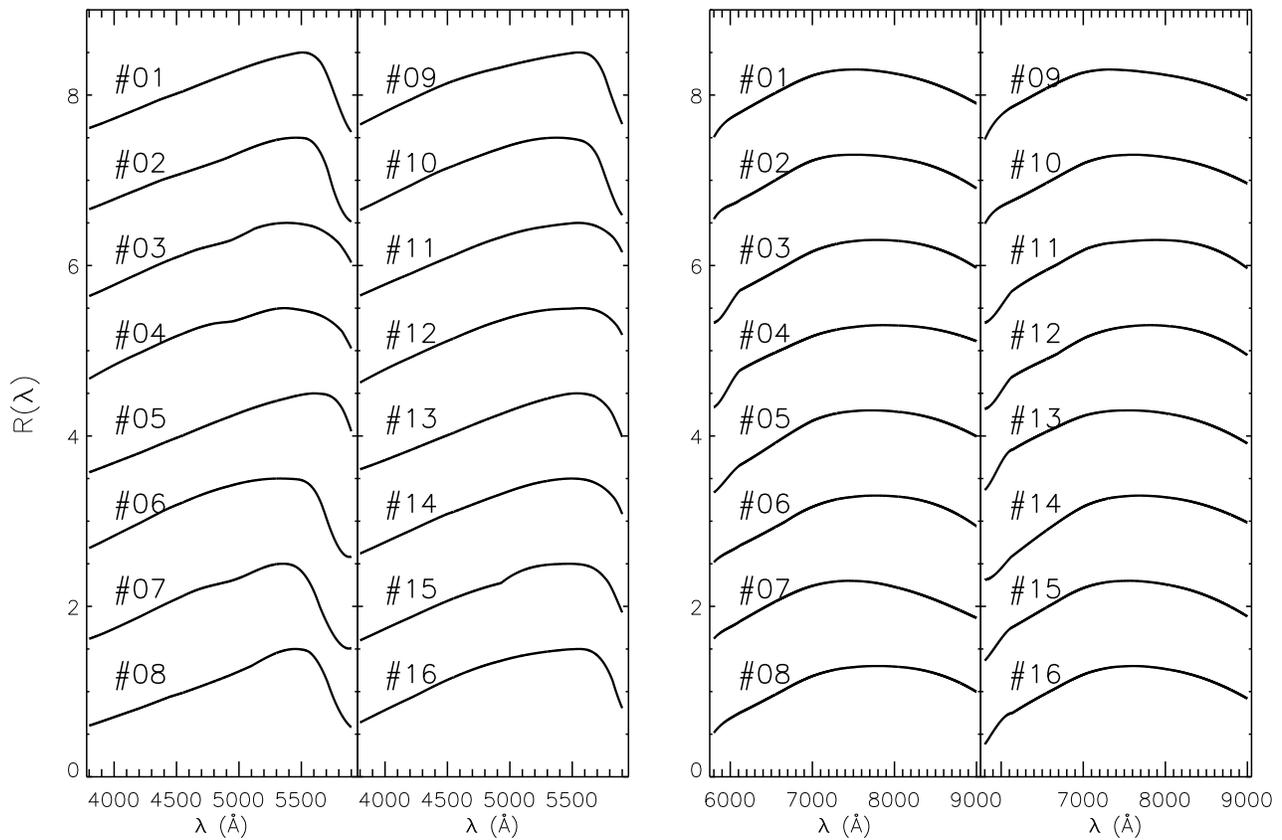}
\caption{Examples of SRCs of the 16 LAMOST spectrographs derived
         for one specific plate (B6212) observed on October 27, 2011, 
         for the blue-arm (left two panels) and red-arm spectra
         (right two panels). The spectrograph numbers are marked.}
\label{Fig4}
\end{figure*}

We have used spectrophotometric flux standard stars  
(Oke 1990; Hamuy et al. 1992, 1994) with minimal spectral features, 
such as DZ white dwarfs, to validate that the SRCs are indeed smooth 
function of wavelength. As the surveys progress, the LAMSOT have 
observed a huge number (on the order of 1 million hitherto) of 
very bright stars of optical magnitudes between 9--14\,mag., 
utilizing observing time of grey to bright lunar conditions. 
Among those bright stars observed, we have found 3 spectrophotometric standard stars, 
Feige\,34, GD\,71 and Hz\,21, that have been observed with the LAMOST.
Fig.\,5 shows the response curve, $R(\lambda)$ [cf. Eq. (1)], 
derived from the LAMOST spectra of the spectrophotometric standard stars.  
The intrinsic flux density as a function of wavelength of those spectrophotometric 
standard stars are retrieved from the ESO websites\footnote{
https://www.eso.org/sci/observing/tools/standards/spectra.html}, 
and the same piecewise low-order polynomials as adopted to fit the SRCs 
yielded by the LSS-GAC standard stars are used to fit the data 
in Fig.\,5. For Feige\,34, residuals caused by the stellar absorption 
lines are seen due to the differences in spectral resolution between the LAMOST 
spectra and the spectra of spectrophotometric standard stars retrieved from the ESO website. 
For GD\,71 and HZ\,21, because of the low SNRs of 
the red-arm spectra, only the results from the blue-arm spectra are shown. 
The GD\,71 and HZ\,21 are among the faintest 
stars observed in the corresponding plates, designed to target 
``very bright'' stars of magnitudes between 9 -- 14\,mag. in 
the optical. As a result, the SNRs of the spectra are 
not optimal. The spectrum of GD\,71 from a single exposure 
has a SNR of about 28 per pixel ($\sim$\,1.07\,{\AA}) 
at 4650\,{\AA} and 5 per pixel ($\sim$\,1.7\,{\AA}) at 7450\,\AA, 
while the corresponding values of the spectrum of HZ\,21 are about 12 and 1, respectively. 
Nevertheless, Fig.\,5 shows that piecewise low-order polynomials 
fit the SRCs well in all cases. 
There are no obvious patterns in the residuals of fit, barring regions strongly 
affected the telluric bands. Note that there is a shallow dip between 
4800 and 5100\,{\AA} in the SRCs deduced from the spectra of 
Feige\,34 and GD\,71, both collected with spectrograph \#3. 
Similar features are found in SRCs derived from 
F-type standard stars observed with spectrographs \#\#3 -- 4, 
\#\#\,7 -- 8, and \#\,15 (Fig.\,4), and consequently probably real. 
\begin{figure*}
\centering
\includegraphics[width=18cm]{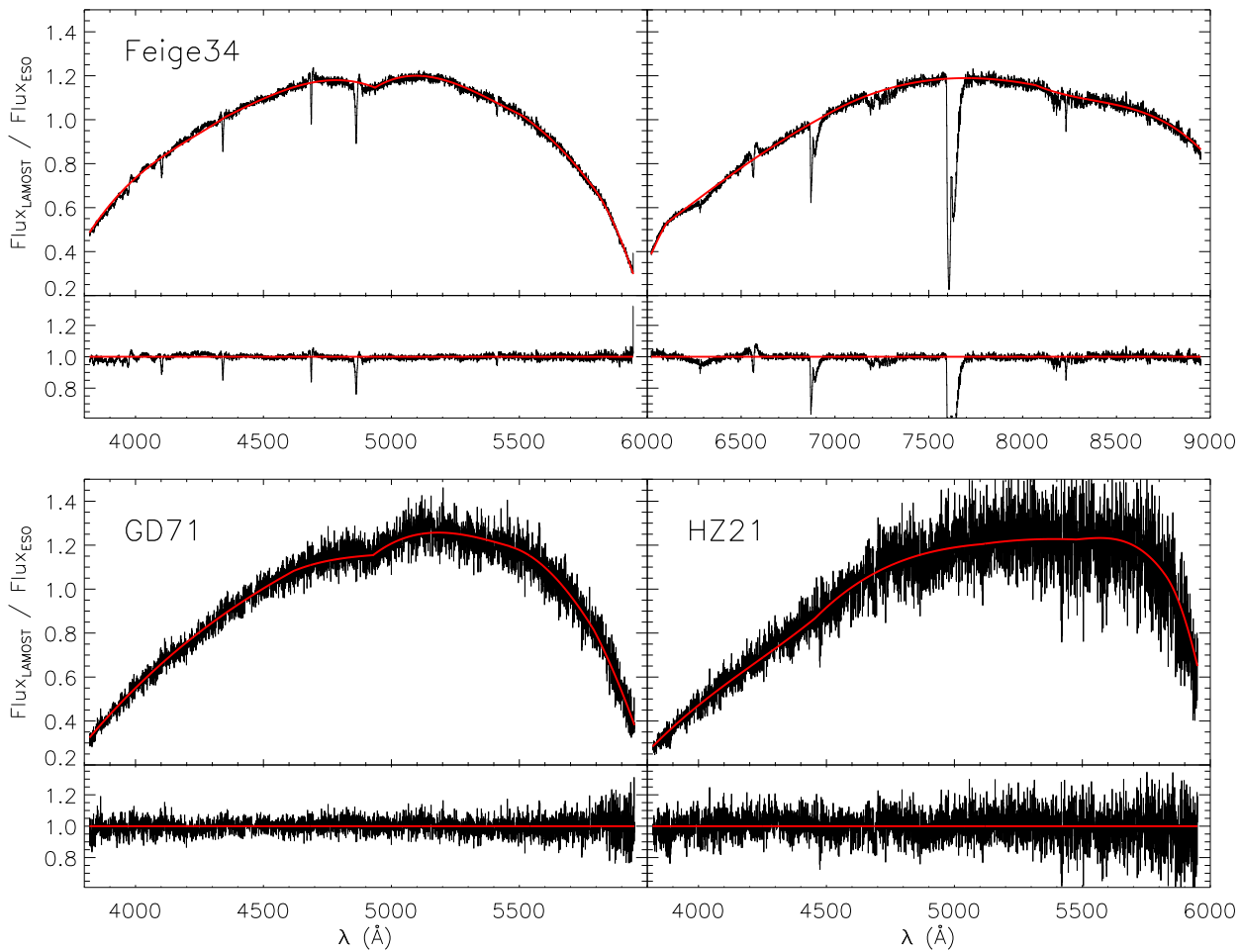} 
\caption{SRCs derived from the LAMOST observation of spectrophotometric
        standard stars Feige\,34, GD\,71 and HZ\,21. The red lines 
        are piecewise polynomial fits to the data. 
        For Feige\,34, the left and right panels are 
        for the blue- and red-arm spectra, respectively. For GD\,71 and HZ\,21, 
        results from the red-arm spectra are not shown due to the poor spectral SNRs. 
        Note that regions affected by the prominent telluric absorption bands 
        have been masked out when fitting the SRCs.
        The bottom of each panel shows the fitting residuals.}
\label{Fig5}
\end{figure*}

\subsection{The final spectra}

To flux-calibrate the spectra, 1-D spectra produced by the LAMOST 2-D pipeline, 
after corrected for the effects of Earth atmospheric extinction, are divided by the SRCs. 
Generally, 2 to 3 consecutive exposures of equal integration time 
are obtained for each LSS-GAC plate. 
To improve the SNR, spectra from the individual exposures are co-added. 
In doing so, a linear algorithm with strict flux conservation is adopted 
to re-bin the spectra of individual exposures to a common wavelength grid. 
The approach, at the cost of some marginal degradation in the spectral resolution, 
is different from the default of the LAMOST 2-D pipeline which uses 
spline fitting to re-bin the spectra. 
Once flux-calibrated, the blue- and red-arm 
spectra are pieced together directly without any scaling or adjustment 
to yield the final spectra. 
Excluding the edges of very low sensitivities, the blue- and red-arm 
spectra are simply averaged in the overlap wavelength region. 
Due to the uncertainties in flat-fielding and sky-subtraction, 
some spectra, in particular those of low SNRs can not be pieced 
together smoothly with the current approach. The problem will be 
further discussed in \S\,{3.2.1}. 
Considering that prominent telluric absorption features, such as 
the Fraunhofer's A-band at 7590\,{\AA} and B-band at 6867\,{\AA} 
are strongly saturated, and signals of any generic stellar features 
embedded in those bands are likely totally lost, we have opted not 
to artificially remove those telluric bands in the final, flux-calibrated 
spectra via SRC fitting. 

\section{Results and discussions}

\subsection{Accuracy of the calibrated spectra}
Our method has been successfully applied to the LSS-GAC survey. 
To investigate the accuracy of spectra thus calibrated, 
the results are closely examined, including a careful comparison 
of the SRCs yielded by the individual flux standard stars in a given 
spectrograph of a given plate, a comparison of the calibrated spectra 
from multi-epoch observations of duplicate targets, a comparison of 
the colours yielded by the calibrated spectra and the photometric 
measurements, as well as comparisons 
of the calibrated spectra with external data, such as  
spectra of spectrophotometric standard stars and spectra collected by the SDSS. 

\subsubsection{SRCs derived from the individual standard stars}
Accurate determination of SRCs is essential for accurate spectral 
flux calibration. In this section, the SRCs derived from the 
individual standard stars observed with a given spectrograph 
are compared to test the precision of SRCs deduced. 
Since, as mentioned in \S\,{1}, we concentrate on the relative rather 
than absolute flux calibration, we examine the dispersion 
of relative variations of SRCs yielded by the individual standard stars for a given spectrograph. 
For a given spectrograph of a specific exposure, SRCs yielded by the 
spectra of the individual standard stars are divided by the final (average) 
SRC. The results are scaled to unity at a specific wavelength $\lambda_2$. 
The standard deviation of values of SRCs thus normalized is then calculated 
at a different wavelength, $\lambda_1$, denoted as $s.d.(\lambda_1)$. 
The latter is adopted as a measure 
of the uncertainties of the shape of SRC derived from the individual 
standard stars for the given spectrograph. Here the subscript `n' 
denotes `normalized'. For the blue-arm spectra, $\lambda_1$ and $\lambda_2$ are 
set to be 4100 and 5600\,{\AA}, respectively, while for the red-arm spectra, 
the corresponding values are 6300 and 8800\,{\AA}. 
As an example, the left panel of Fig.\,6 plots SRCs as well as 
$R(\lambda)/R(\lambda_2)$ as a function of wavelength $\lambda$ yielded by 
individual exposures of individual standard stars observed with 
spectrograph \#\,6 of plate `20111028-GAC060N28B1'. The Figure shows 
that although the absolute values of SRCs yielded by the 
individual standards may differ by up to a factor of 2 -- 3, 
their shape between 4100 and 5600\,{\AA} vary by less 
than 5 per cent. The right panel of Fig.\,6 shows the distribution of 
uncertainties of the relative SRCs, quantified by $s.d.(\lambda_1)$, 
for all the spectrographs of all 602 plates collected by
June, 2013 that are included in the current analysis. It indicates 
that for the majority of spectrographs (67 per cent for the blue and 84 per cent for the red-arm), 
the uncertainties in terms of the relative SRCs are smaller than 10 per cent 
for both the blue- and red-arm spectra.  
Note that almost all spectrographs of large uncertainties 
in the relative SRCs belong to VB plates. 
If we exclude VB plates, 88 and 93 percent of the spectrographs 
have uncertainties less than 10 per cent 
for the blue- and red-arm spectra, respectively.   
The large uncertainties in both interstellar extinction corrections 
and sky subtractions for the VB plates, observed under bright lunar 
conditions and other unfavorable observing conditions, may have contributed 
to the large SRC uncertainties for those plates. 
Since we have adopted the average of SRCs yielded by the individual standard 
stars as the final SRC, we expect that the uncertainties of the latter 
are smaller by a factor of 2 -- 3 given that the majority of 
spectrographs have 5 -- 10 standard stars. 

\begin{figure*}
\centering
\includegraphics[width=18cm]{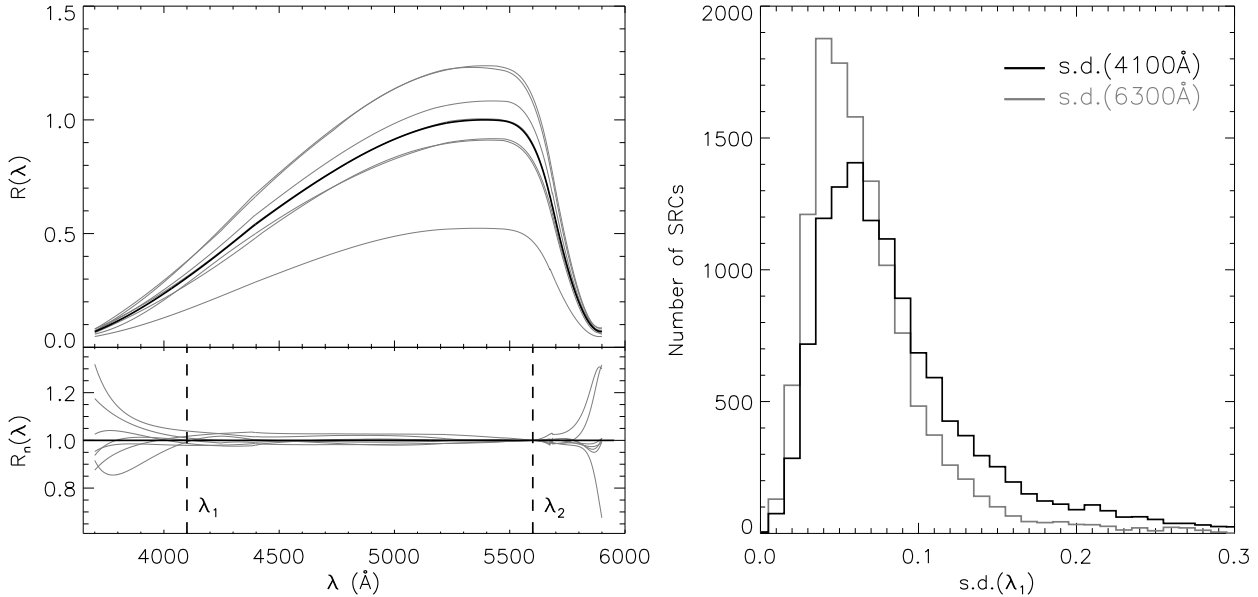}
\caption{{\em Upper-left:}\, $R$($\lambda$) derived from the 
         individual standard stars (grey) and their SNR-weighted mean 
         (black) for a given spectrograph of a given exposure. 
         {\em Lower-left:}\, SRCs from the individual standards are 
         divided by the SNR-weighted-mean and then normalized to unity 
         at wavelength $\lambda_2$, denoted as $R_{\rm n}(\lambda)$. 
         The vertical dashed line marked as $\lambda_2$ indicates the wavelength 
         of normalization, while the dashed line marked as $\lambda_1$ indicates 
         the wavelength where the standard deviation of $R_{\rm n}(\lambda_1)$, 
         $s.d.(\lambda_1)$, is calculated. For the blue-arm spectra, 
         $\lambda_1 = 4100$\,{\AA} and $\lambda_2 = 5600$\,{\AA}. 
         For the red-arm spectra, the corresponding values are 6300 and 8800\,{\AA}, respectively. 
         {\em Right}\,: Distribution of the standard deviations of $R_{\rm n}(\lambda_1)$ 
         yielded by all the spectrographs of all exposures included in the current analysis.}
\label{Fig6}
\end{figure*}

\subsubsection{Comparison with photometry}
To test the (relative) accuracy of our flux-calibration algorithm, we 
convolve the flux-calibrated spectra with the SDSS $g$, $r$ and 
$i$-band transmission curves and compare the resultant 
colours with the XSTPS-GAC photometry, which is globally calibrated 
to the SDSS photometric system to an accuracy of 0.02 -- 0.03\,mag (Liu et al. 2014). 
For this purpose, we select the LSS-GAC plates collected in 
the Pilot Surveys (October 2011 -- June 2012) and the first year 
of the Regular Surveys (October 2012 -- June 2013). 
We require that the stars are fainter than 13.0\,mag in all $g$, $r$ and $i$ bands 
to ensure that the stars are not saturated in the XSTPS-GAC survey, 
and that the spectra 
have SNRs higher than 20 to minimize the uncertainties caused by poor 
sky-subtraction. Spectral plates observed with lunar distances 
smaller than 60\,deg., as well as plates that lack enough 
number of standards of sufficient SNRs and are thus calibrated 
using SRCs deduced from other plates, are excluded. 

Distribution of the differences of $(g - r)$ and $(g - i)$ colours 
yielded by the flux-calibrated spectra and those from the XSTPS-GAC 
photometry are shown in Fig.\,7. For $(g - r)$, with a difference of 
$0.02\pm 0.07$\,mag., the results agree extremely well. 
For $(g - i)$, colours yielded 
by the spectra are on average 0.04\,mag bluer than the photometric values. 
The small systematic offset is due to the fact that we have 
opted not to correct for the telluric features in the 
spectra, most notably the Fraunhofer A-band at 7590\,{\AA} and B-band at 6867\,{\AA} 
in the wavelength range of photometric $i$-band (cf. \S\,{2.5}).

\begin{figure}
\centering
\includegraphics[width=8cm]{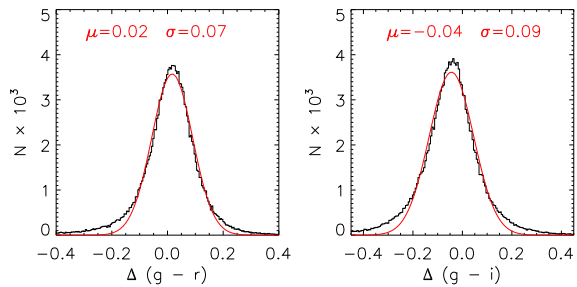}
\caption{Histograms of differences of $(g-r)$ (left) and $(g-i)$ (right) 
         colours yielded by the flux-calibrated LSS-GAC spectra and those given 
         by the XSTPS-GAC photometric survey, for a sample of 158,734 stars. 
         Also over-plotted are Gaussian fits to the distributions, with the 
         mean and dispersion of the Gaussian marked.} 
\label{Fig7}
\end{figure}

\subsubsection{Comparison of multi-epoch observations}
In the LSS-GAC survey, $\sim$\,23 per cent of the sample stars are targeted 
more than once due to the overlapping of FoV of the adjacent plates (Liu et al. 2014). 
As a check of the robustness of our flux calibration algorithm, 
we compare the spectra of common objects acquired in different plates, 
usually taken on different nights. 
To minimize uncertainties of sky subtraction, only 
spectra 
with SNRs higher than 30 per pixel at 4650\,\AA\, and observed with 
lunar angular distances larger than 60\,deg. are used for the comparison. 
We further require that the difference between the spectral and 
photometric $(g-r)$ colours is less than 0.16\,mag, i.e. within $\sim$\,2$\sigma$ 
of the distribution shown in Fig.\,7, to exclude spectra with 
poor alignment of the blue- and red-arm spectra, caused by for example, 
poor flat-fielding.  
The selection yields 15,019 pairs of spectra of common targets. 
The two spectra of each pair are ratioed and then scaled to 
a mean value of unity for the wavelength range 4060 -- 4080\,\AA, 
a relatively clean region devoid of prominent spectral features. 
The results are plotted in Fig.\,8. The ratios yield an average 
that is almost constant for the whole spectral wavelength coverage, 
with a standard deviation of 2 and 15 per cent at 4070 and 9000\,{\AA}, respectively. 
The results show that our flux-calibration algorithm has achieved a 
precision of $\sim$\,13 per cent between 4100 and 9000\,{\AA}. In fact, since 
both spectra of a pair contribute to the dispersion, the underlining 
precision of a given spectrum should be slightly better, probably at 
the level of 10 per cent. The rapid increase 
of scatter at shorter wavelengths is mainly caused by the 
rapid decline of the instrumental throughput and thus the limited 
SNRs of spectra at those wavelengths. 

\begin{figure}
\centering
\includegraphics[width=8cm]{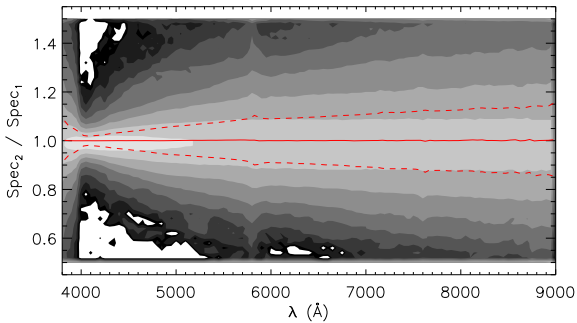}
\caption{Distribution of ratios of spectral pairs of common objects 
         secured at different epochs. 
         The grey contours are shown on a logarithmic scale, 
         and represent the distribution of values of ratios of 15,019 spectra pairs. 
         The mean and standard deviation of the ratios as a function of wavelength are shown by the 
         solid and dashed curves, respectively.} 
\label{Fig8}
\end{figure}

\subsubsection{Comparison with spectrophotometric standards}
In this subsection, we compare the absolute spectral flux densities 
yielded by the calibrated LAMOST spectra directly with values retrieved 
from the ESO website (denoted as `ESO flux density' hereafter) 
and those from the CALSPEC database of $Hubble$ $Space$ $Telescope$ $(HST)$ 
standards \footnote{http://www.stsci.edu/hst/observatory/crds/calspec.html 
(Bohlin et al. 2014; denoted as `CALSPEC flux density' hereafter)} 
for the spectrophotometric standard stars Feige\,34, GD\,71 and HZ\,21. 
The results are shown in Fig.\,9. For GD\,71 and HZ\,21, only results 
from the blue-arm spectra are shown. The red-arm spectra have too low SNRs ($<5$) 
to be useful. 
The Figure indicates that although the absolute flux densities given by 
the calibrated LAMOST spectra are a factor of 1 -- 3 lower than the ESO 
values, their ratios, i.e. the shapes of the SEDs agree well. 
Comparison with the CALSPEC database yields similar results.
For Feige\,34, except for regions affected by the telluric bands 
(e.g. the O$_2$ a-band at $\sim$\,6280\,{\AA}, the O$_2$ B-band at $\sim$\,6870\,{\AA}, 
the H$_2$O band at $\sim$\,7200\,{\AA}, the O$_2$ A-band at $\sim$\,7600\,{\AA} 
and the O$_2$ z-band at $\sim$\,8220\,{\AA}) 
and by the broad stellar hydrogen and helium absorption lines, 
the shapes of SEDs given by the LAMOST and 
ESO spectra agree within 5 per cent for the whole wavelength range. 
Similar result is achieved for HZ\,21, except near the red end of 
the wavelength coverage. For GD\,71, the shape of SED given by the LAMOST 
spectrum deviates from that of the ESO 
spectrum by up to 20 per cent between 4000 and 5000\,{\AA}. 
The cause of the large deviation is not clear. 
We suspect that some fibers may have poorly flat-fielded. 
Specifically, the spectral response of some fibers may have varied 
between the observation of target plate and that of 
the twilight, the latter is used to flat-field the fibers in the current 
implementation of 2-D pipeline (\S\,{3.2.1}).  
Another possibility is that the extinction towards some of the selected 
standard stars are significantly ($>0.1$\,mag) over-estimated for some 
unknown reasons (e.g. contaminations from a binary companion). In fact, 
for this particular observation of GD\,71, a subset of 4 (out of 10) 
standard stars yield a SRC that recovers the ESO flux density. 
The spectrum of GD\,71 calibrated using the SRC deduced from this subset 
of standard stars is shown in grey in Fig.\,9. 
\begin{figure}
\centering
\includegraphics[width=8cm]{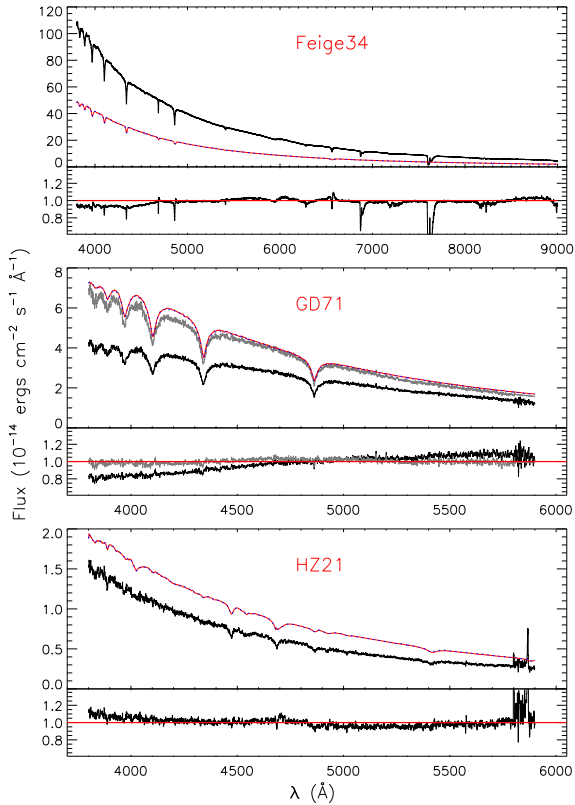}
\caption{Comparison of calibrated LAMOST spectra (black) with spectra retrieved from 
         the ESO website (red lines) and from the CALSPEC database 
         (blue dashed lines; almost indistinguishable from the red lines) 
         for flux standard stars Feige\,34 (top), GD\,71 (middle) and HZ\,21 (bottom). 
         In each panel, the upper part shows the absolute flux density as 
         a function of wavelength, whereas the lower 
         part shows the ratios of the LAMOST and ESO flux density,  
         normalized to the median value of the ratios. Note for GD\,71 and HZ\,21, 
         the red-arm spectra are not shown due to poor SNR. 
         Spectrum of GD\,71 processed with the SRC yielded by 
         a sub-set of 4 selected standards is shown in grey (see text for detail).}
\label{Fig9}
\end{figure}

\subsubsection{Comparison with SDSS spectra}
We have made an extensive comparison of flux-calibrated LSS-GAC
with those of SDSS for common objects, and found 
that for stars of high Galactic latitudes suffering from low interstellar 
extinction, they match with each other quite well, with differences
at the level of only a few percent for the whole wavelength range, 
provided the SNRs of both sets of spectra are higher than 30.
For stars of low Galactic latitudes and suffering from high extinction,
large deviations between them are seen for some sources. 
Some examples are illustrated in Fig.\,10.
The Figure shows that for stars suffering from low extinction, the SEDs yielded
by the LSS-GAC spectra match quite well with those of SDSS, except for
regions strongly affected by the telluric bands which have not been 
removed from the LSS-GAC spectra.
For stars suffering high extinction, the SEDs yielded by SDSS 
are typically redder than those of LSS-GAC. What's more, the
SDSS spectra may show strange `S'-shaped SEDs.
We believe that this is likely caused by unrealistic SRCs of the SDSS 
for high extinction fields as a consequence that the effects of the 
interstellar extinction have not been properly accounted for by the SDSS 
pipeline. The SDSS pipeline uses the SFD98 extinction map 
to correct for the effects of the interstellar extinction. 
However, the SFD98 map gives the total extinction integrated along 
a given line-of-sight and generally over-estimates the extinction for 
a disc star in that direction of line-of-sight. 
The redder SEDs yielded by SDSS spectra compared with those of LSS-GAC 
for stars suffering from large amounts of extinction can be explained if 
the SDSS pipeline has over-estimated the extinction of flux-calibration standard
stars. The different extinction laws adopted by the 
LSS-GAC and SDSS pipelines may also play a role. The SDSS adopts
the O'Donnell's (1994) extinction law, while we use that of Fitzpatrick.
It is possible that the S-shaped SEDs seen in some of the SDSS spectra 
is a defect caused by the inadequacy of the 
O'Donnell extinction law, and the defect becomes more severe for stars 
suffering from higher extinction. 
In fact, it has been argued that the Fitzpatrick law fits 
the data better than the O'Donnell's (Schlafly et al. 2010; 
Berry et al. 2012; Yuan et al. 2013). Note that in Fig.\,10,  
some spikes caused by cosmic rays that have not been cleaned 
can be seen in some of the LSS-GAC spectra plotted (cf. XLYH14). 
\begin{figure*}
\centering
\includegraphics[width=18cm]{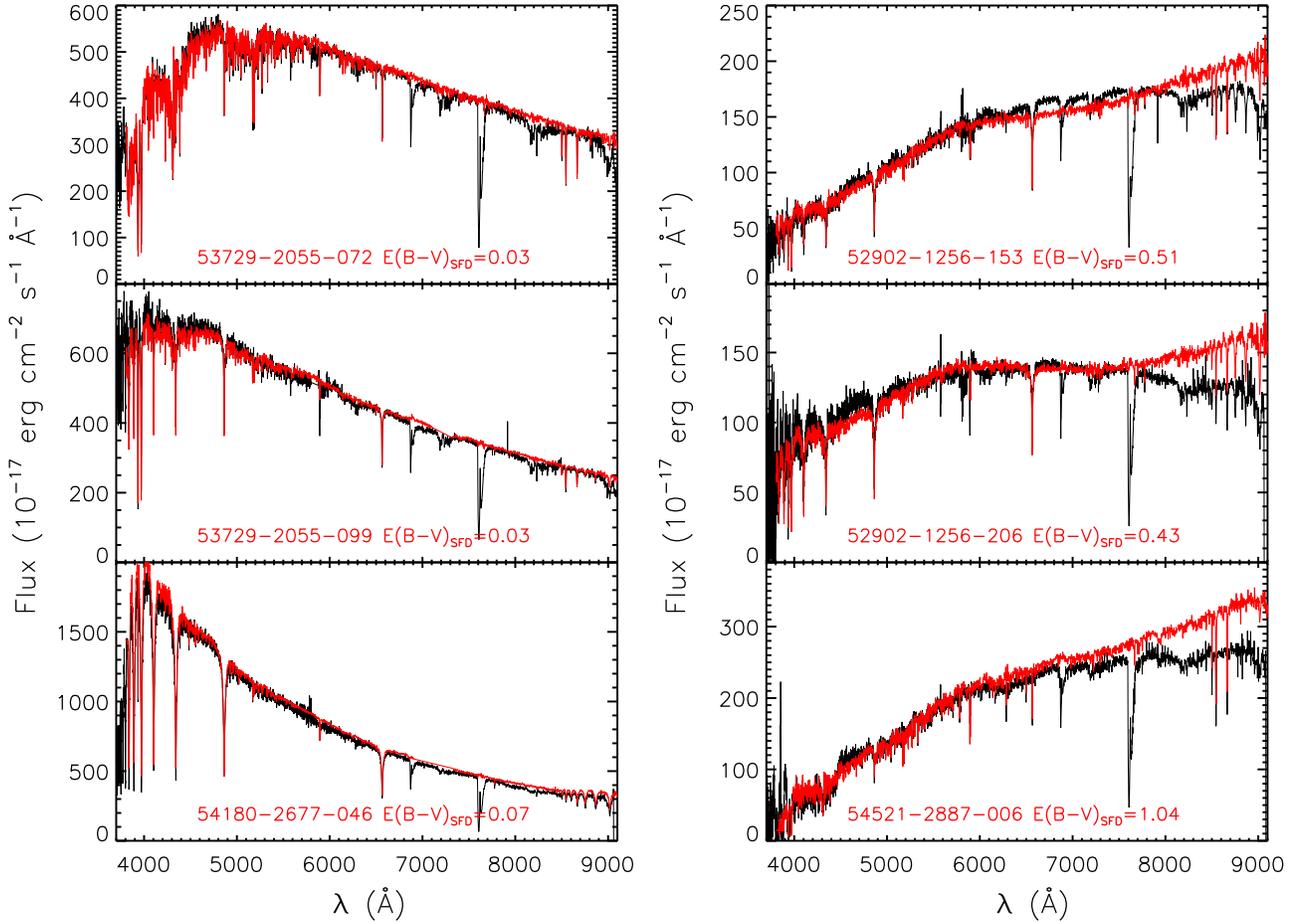}
\caption{Comparison of LSS-GAC (black) and SDSS spectra (red). 
         The left panel shows stars suffering from low reddening, 
         whereas those plotted in the right panel suffer from significant 
         amount of reddening. The SDSS object ID, in the format
         of MJD-plateid-fiber, as well as the $E(B-V)$ from
         the SFD98 reddening map, are marked in each panel in red. 
         Note that the telluric bands have 
         not been removed from the LSS-GAC spectra.} 
\label{Fig10}
\end{figure*}

\subsection{Error sources of the flux calibration}
There are several obstacles that prevent one from obtaining 
accurately calibrated spectra with the LAMOST-like multiplex telescopes
that have a very wide FoV, 
including obtaining homogeneous and stable flat fields,
accurate sky and other background light (e.g. scattered light) subtraction 
in 2 dimensions, and accurate fiber positioning and so on. In this subsection, 
we discuss the potential 
effects on the flux calibration from those difficulties. 

\subsubsection{Flat fielding and sky subtraction}
To ensure the fiber system work stably is an 
important issue for the LAMOST. There is evidence that 
the SRCs of the individual fibers vary from observation of 
one plate to another, probably caused by the changes in stress 
of the fibers after repositioning. The variations amount   
to 10 per cent across the whole wavelength range covered 
by the LAMOST spectra (Chen J.-J., private communication). 
It suggests that fiber flat fields derived from twilight exposures  
may not be suitable for flat-fielding the science exposures obtained during the night.
The variations will in no doubt have an impact on sky subtraction and flux
calibration, introducing uncertainties to the final calibrated spectra.
Uncertainties introduced by such variations of fiber flat fields 
do not depend on the spectral SNRs. 
As a consequence, even spectra of very high SNRs may have 
an incorrect shape of SED. 
Attempts to characterize and correct for such variations of 
fiber flat fields is under way (Chen J.-J., private communication). 

To minimize potential errors introduced by poor sky subtraction, 
the current implementation of LAMOST 2-D pipeline (v2.6) scales 
the sky spectrum to be subtracted from the target spectra by the 
measured fluxes of sky emission lines, assuming that latter are  
homogeneous across the FoV of individual spectrographs (about 1\,deg.). 
The difficulty is that, given that the continuum sky background 
and sky emission lines originate from very different sources and 
excited by different mechanisms, their emission levels are unlikely 
scale linearly with each other. In fact, even amongst the sky emission 
lines, lines from different species (such as atomic [O I] and molecular OH) 
may have quite different behavior in terms of their temporal and 
spatial variations. Scaling the sky spectra by the measured fluxes 
of sky emission lines risk subtracting incorrect level of continuum 
sky background. 
To ensure that the blue- and red-arm spectra join smoothly, 
the default flux calibration algorithm in the LAMOST 2-D pipeline 
forces the blue spectrum between 5295 -- 5700\,{\AA}
and the red spectrum between 6040 -- 6530\,{\AA} to line with each other 
by shifting and scaling the red-arm spectrum in a sort of arbitrary way. 
We have opted not to perform such shifting or scaling. 
As a consequence, the blue- and red-arm spectra of some of the spectra 
processed with our own pipeline do not join smoothly. For spectra of 
SNRs higher than 10, the discontinuity, if any, amounts to only a few 
per cent in most cases.

\subsubsection{Spectral SNRs}
\begin{figure}
\centering
\includegraphics[width=8cm]{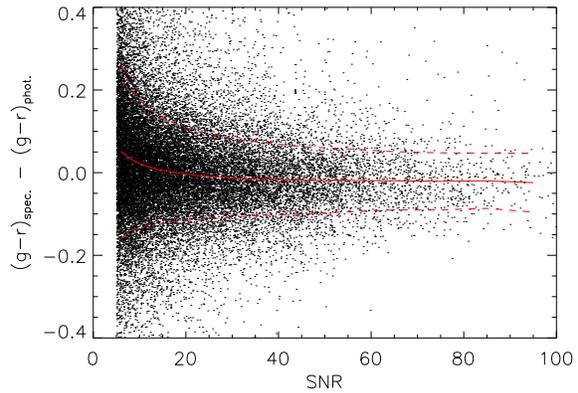}
\caption{Difference of spectral and photometric $(g-r)$ colours 
         as a function of the spectral SNR. 
         The solid and dashed lines in red delineate the mean and 
         standard deviation of the differences. 
         For clarity, only one in ten data points are shown.}
\label{Fig11}
\end{figure}

Spectral SNRs obviously have a large impact on the accuracy 
by which the spectra can be calibrated. 
To test how the SEDs of LSS-GAC targets are affected by 
limited SNRs, we compare the spectral and photometric $(g-r)$ 
colours as a function of the spectral SNR.
The results are plotted in Fig.\,11. The Figure shows that 
at high SNRs, the two agree well, with a mean difference 
of 0.01 -- 0.02\,mag and no systematic trend. At SNRs lower 
than about 10, the discrepancies increase rapidly, along with 
some systematic differences. Clearly, poor sky subtraction 
dominates at such low SNRs. Note that even at SNRs higher than 
40, there is still a significant scatter of 0.06\,mag 
between the spectral and photometric colours. This is likely 
caused uncertainties in the SRCs as well as fiber flat fielding. 

Poor spectral SNRs may also have a large impact on the accuracy 
of flux calibration due to the restrictive number of quality flux 
calibration standards available for a given spectrograph or plate.
As described in \S\,{2.3}, about 10 per cent plates observed hitherto 
fail to yield sufficient numbers of standard stars of good SNRs. 
Those plates are processed with SRCs derived from other plates of 
better quality. Flux calibration of those plates is thus unreliable. 
They are mostly M or F plates, or B and VB plates observed under 
poor conditions. Many of them are collected during the Pilot Surveys. 
With better observational planning and more strict quality control, 
the number of such plates are much reduced as the Regular Surveys progress.

\subsubsection{Stellar atmospheric parameters}
We show in \S\,{2.2} that for flux standard stars of 
5750\,$\leq$\,$T_{\rm eff}$\,$\leq$\,6750\,K, an error of 150\,K 
in $T_{\rm eff}$ can lead a maximum uncertainty of 12 per cent in 
the shape of the stellar SED and thus the shape of SRC derived from it. 
Uncertainties caused by errors in log\,$g$ is negligible, about 
1 per cent at most for the whole wavelength range concerned for 
an estimated error of 0.25\,dex in log\,$g$. 
The metallicity mainly affect the blue-arm spectra at short wavelengths 
($\lambda<4500$\,\AA). A difference of 0.2\,dex in 
[Fe/H] can change the SED shape between 3800 and 4500\,{\AA} 
by approximately 3 per cent, while the effects at $\lambda>4500$\,{\AA} 
are only marginally. Since the SRC of each spectrograph is 
derived using several (up to 10) standard stars, the impact of 
uncertainties in stellar atmospheric parameters of individual  
flux standard stars on the final SRC deduced are much reduced 
and expected to be at the level of a few per cent at most.  

\subsubsection{Extinction towards the flux calibration standard stars}
Comparison with the SFD98 reddening map for 
high Galactic latitude stars show that typical uncertainties 
of $E(B-V)$ derived with the method adopted in the current work 
is about 0.04\,mag (Yuan et al. 2014b submitted). 
An uncertainty in $E(B-V)$ of this level can lead to an error of 
 $\sim$\,10 per cent maximum in the deduced SRCs. Also it is quite 
possible that some of the selected flux calibration standard stars 
are actually binaries or multiple stellar systems. In such case, 
values of $E(B-V)$ estimated may suffer from large errors. 
The effects of uncertainties in estimates of reddening may be 
more severe for VB plates for which values of reddening for individual 
targets are estimated using the 2MASS $J$, $H$, $K_{\rm s}$ bands only. 
Currently, there is no good solution to this problem. However, 
this effect is possibly insignificant. A comparison with the SFD98 
reddening map shows no clear patterns of errors for values of extinction 
estimated based on the near IR photometry alone.  

In addition to uncertainties in $E(B-V)$, the variations of 
extinction law, or specifically $R_V$ in different Galactic 
environments will also introduce additional uncertainties 
to the flux calibration. Although it is generally accepted 
that $R_V=3.1$ is a good value to use for the general diffuse 
interstellar medium, moderate variations in $R_V$ have been 
found in the Galactic disk (cf. Fitzpatrick \& Massa 2007, and reference therein). 
By comparing the multi-band photometric colours with synthetic 
values from stellar atmospheric models, we find $R_V$ of 3.15 
for all selected flux standard stars used to calibrate all plates 
collected hitherto, along with a standard deviation of 0.25. 
The result is well consistent with that of Fitzpatrick \& Massa (2007). 
A scatter of $0.25$ in $R_V$ can lead to an uncertainty of 10 
per cent in the SRCs deduced. 
However, not all scatters seen in $R_V$ are real, a significant 
portion of the variations is probably caused by measurement uncertainties. 
Again, since more than one flux calibration standard stars are used 
for a given spectrograph, we expect that uncertainties of SRCs introduced 
by errors in extinction corrections for the flux calibration 
standard stars are probably on the level of several to ten per cent in general. 

\subsubsection{Prominent absorption features in spectra of the standard stars}

There are several prominent stellar atmospheric absorption
features in the spectra of F-type standard stars, such as the 
hydrogen Balmer lines, the CH G-band at 4314\,{\AA} and a number 
of strong lines from metal Fe, Ti and Mg between 5100 -- 5300\,{\AA}. 
Although spectral regions affected by those prominent features 
have been masked out in fitting the SRCs, their effects cannot 
be fully neutralized, in particular at wavelengths $\lambda<4000$\,{\AA}, 
where the spectra are plagued by the series of high order hydrogen 
Balmer lines as well as the Balmer discontinuity.
In addition, the instrumental sensitivity drops rapidly to 
very low at such short wavelengths, making the SRCs quite 
uncertain in this wavelength regime. The SRCs at $\lambda\sim$\,3800\,{\AA} 
may be uncertain by as much as 20 -- 40 per cent, or more in some extreme cases. 
More observations of featureless spectrophotometric standard stars, 
such as DZ white dwarfs, may help characterize the shape of SRC in 
this near ultraviolet wavelength regime. 

\subsubsection{Other error sources}
In addition to sources of error discussed above, some other factors 
may also affect the robustness of SRCs derived, including the 
differential refraction of the Earth atmosphere, uncertainties 
in fiber positioning. 
Those effects are however difficult to characterize and may well 
vary from plate to plate. Careful observational planning and strict 
quality control is the key to ensure data quality and the robustness 
of calibration. 

\subsection{Temporal variations of the shape of SRC}
\begin{figure*}
\centering
\includegraphics[width=18cm]{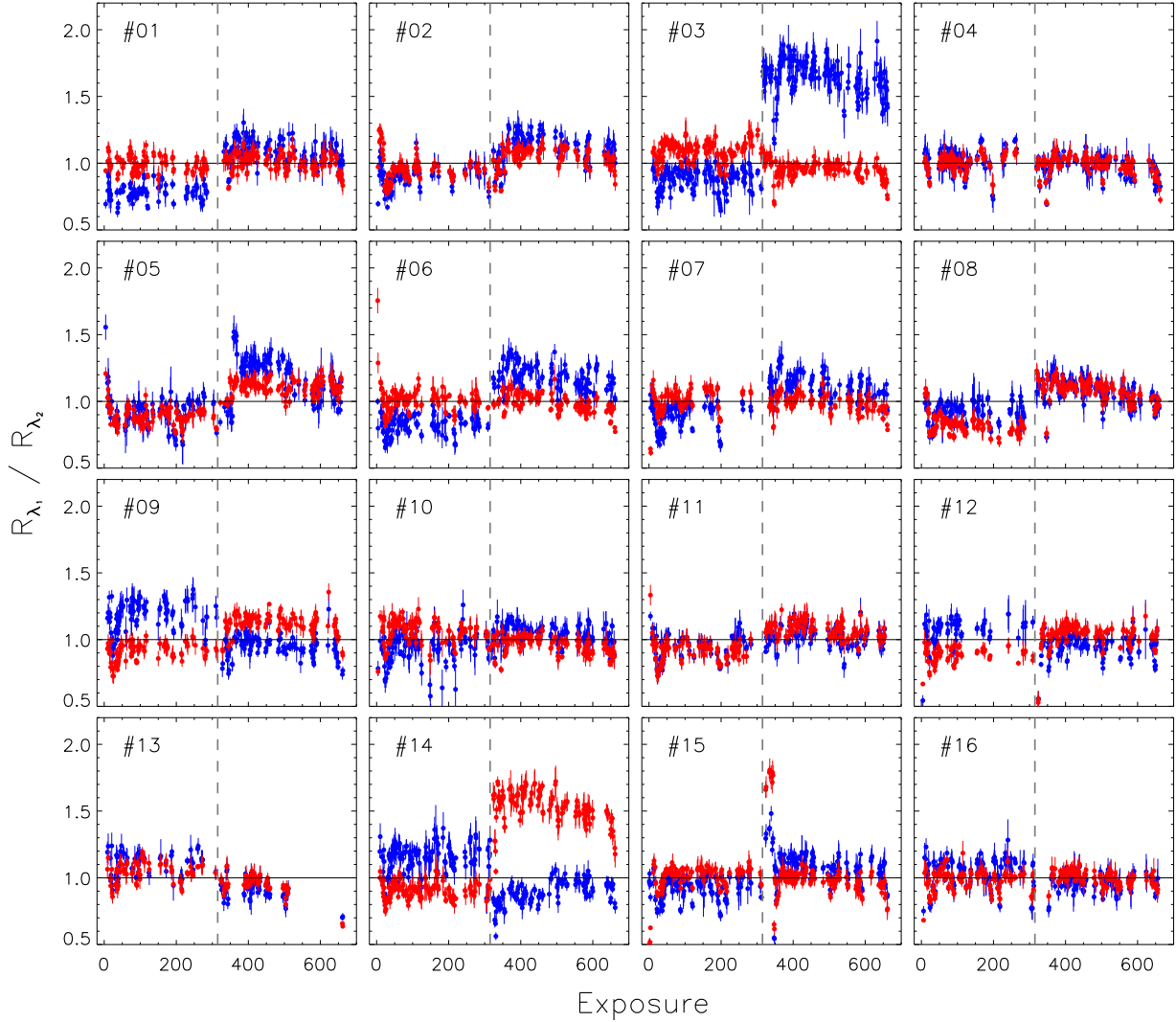}
\caption{Ratios of SRC values at two specific wavelengths, for SRCs 
         deduced from individual exposures for spectrograph \#6, 
         in the sequence of time of observation.
         For the blue-arm SRCs (upper),
         $\lambda_1 = 4100$\,{\AA}, $\lambda_2 = 5600$\,{\AA}.
         For the red-arm SRCs (lower), $\lambda_1 = 6300$\,{\AA},
         $\lambda_2 = 8800$\,{\AA}.
         The error bars denote the standard deviations of SRCs
         yielded by the individual standard stars.
         The spectrograph numbers are labeled. 
         The vertical grey dashed lines denote the completion
         of the Pilot Surveys and the initiation of the Regular Surveys.}
\label{Fig12}
\end{figure*}
In this subsection, we compare the SRCs deduced from individual 
observations (exposures) and investigate their possible temporal variations. 
In Fig.\,12, we plot the ratios of SRC values at two specific 
wavelengths, $R(\lambda_1)/R(\lambda_2)$, in sequence of exposures, 
for the 16 spectrographs of LAMOST. For the blue-arm SRCs, 
$\lambda_1 =  4100$\,{\AA} and $\lambda_2 = 5600$\,{\AA}, 
while for the red-arm SRCs, the corresponding 
values are 6300 and 8800\,{\AA}, respectively. 
To minimize uncertainties, only SRCs of errors in the shape 
less than 0.08 for both the blue and red arms, and yield 
stellar spectral $(g - r)$ colours of the standard stars agree 
with photometric values within 0.05\,mag, are shown in the plot. 
The requirements exclude most VB plates -- standard stars selected 
for those plates are saturated in the XSTPS-GAC survey and thus 
do not have photometric colours. As a result, not all spectrographs plotted 
in Fig.\,12 have the same number of data points.  
Fig.\,12 shows large jumps and scatters. Some jumps are 
clearly related to the instrumental maintenance, adjustment and optimization. 
For example, during the summer of 2012, the collimators 
of several spectrographs were re-coated. As a result, at the 
onset of the Regular Surveys, one sees a significant upward jump 
in $R(\lambda_1)/R(\lambda_2)$ for those spectrographs, in particular for the blue-arms 
of spectrographs \#\#1 -- 3, \#\#5 -- 8 and \#15, indicating significant 
improvement in sensitivity at short wavelengths.

\begin{figure}
\centering
\includegraphics[width=8cm]{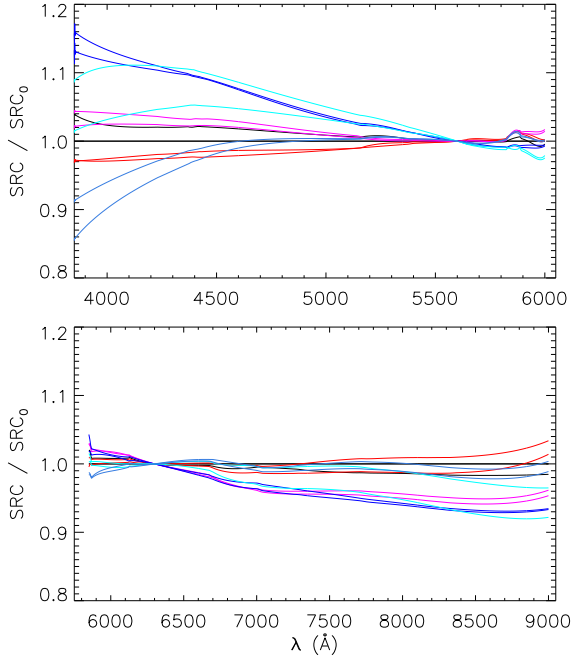}
\caption{SRCs of spectrograph \#5 deduced from the individual exposures 
         taken on 2012 January 12, divided by the SRC yielded
         by the very first exposure of that night. Lines of the same colour represent
         results from the individual (consecutive) exposures of
         the same plate, whereas lines of different colours
         represent results from different plates.}
\label{Fig13}
\end{figure}

Apart from those large jumps in sensitivity, Fig.\,12 also 
reveals significant plate to plate variations 
in the shape of SRCs, at the level of 30 per cent or more. The shape of 
SRCs is found to vary not only on different nights, but also 
in a given night. To illustrate this, we plot in 
Fig.\,13 all SRCs deduced from individual exposures taken 
on January 12, 2011 for spectrograph \#\,5, divided by the SRC 
yielded by the very first exposure 
of that night. In the plot, different colours indicate SRCs from 
different plates (FoV's), whereas lines of the same colour indicates 
SRCs from the individual (consecutive) exposures of the same plate. 
Variations up to $\sim$\,30 per cent are seen in the blue. 
For a given plate, SRCs deduced for the individual (consecutive) 
exposures are generally in good agreement, typically within a couple 
of per cent. 

\begin{figure}
\centering
\includegraphics[width=8cm]{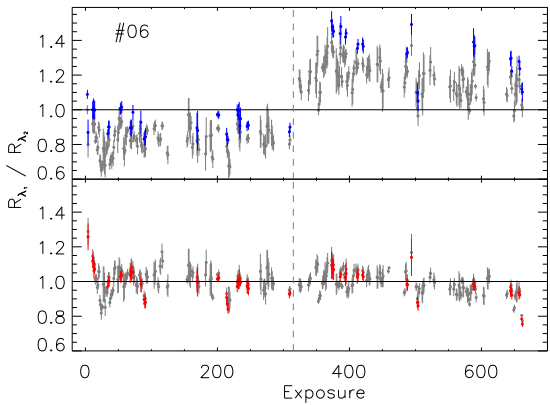}
\caption{Same as Fig.\,12 but for spectrograph \#6 only. 
         Data points in colour are from high Galactic 
         latitude fields ($|b|>25$\,deg.) where the median value of  
         $E(B-V)$ of the individual standard stars used to derive the 
         SRC is smaller than 0.08\,mag.}
\label{Fig14}
\end{figure}

To ensure that the variations in the shape of SRCs seen in Fig.\,12 
are not artifacts induced by improper extinction corrections for flux 
calibration standard stars in the Galactic disc, the same data for 
spectrograph \#{6} are re-plotted in Fig.\,14, highlighting data points 
from high Galactic latitudes. 
In the plot, the data points in colour are from high Galactic latitudes 
fields ($|b| > 25$\,deg.) for which the median values of $E(B - V)$ of 
individual flux-calibration standard stars, as estimated by comparing 
the measured and synthetic photometric colours as well as given by the 
SFD98 extinction map, are both smaller than 0.08\,mag.
The Figure clearly shows that the variations cannot be due to 
uncertainties in reddening corrections. 
As a final check that the variations are not due to some hidden 
errors in our flux calibration procedure, we have directly compared 
the SEDs of uncalibrated spectra yielded by 
the LAMOST 2-D pipeline, of stars observed at multi-epochs. 
Again the comparison confirms that the variations of SRCs are real. 
The large variations in the shape of SRCs, from plate to plate, 
suggest that deriving SRCs plate by plate 
is essential for accurate flux calibration of the LAMOST spectra.

The causes of such plate to plate variations in the shape of SRCs are not clear. 
They could be caused by stress or strain induced variations of fiber 
spectral response when the telescope moves from one plate (field) to 
another and the fibers reposition, as mentioned in \S{3.2.1}, and/or 
by colour-dependent variations of image quality cross the FoV as the 
telescope tracks during the exposures. 
Another possibility is the change of the Earth atmospheric 
extinction curve during the night.  
Due to the increasing air pollution at Xing Long Station, in particular 
in winter, the local atmospheric reddening and extinction are known 
to exhibit erratic variations at short time scale. 
Further studies are needed however to clarify the situation.

\section{Summary} 

We have presented the relative flux calibration for the LSS-GAC. 
Stars of $T_{\rm eff}$ between 5750 and 6750\,K as yielded by the LSP3, 
most of which are F-types, are selected as flux calibration standard stars. 
For the majority of LSS-GAC plates and spectrographs, more than 
4 standard stars can be selected out. The SRCs are derived by 
fitting with low-order piecewise polynomials on the ratios of 
the observed spectra of the standard stars and the synthetic ones 
that have the same stellar atmospheric parameters as the standard stars, 
after reddened the latter using values of the interstellar reddening 
$E(B-V)$ derived by comparing the observed and synthetic colours, 
assuming a $R_V=3.1$ Fitzpatrick extinction law. 
For plates that not enough standard stars can be selected out, 
the SRCs deduced form other plates, usually observed on the same night, 
are used to process the spectra. 
Spectra from the individual consecutive exposures of the same plate are co-added 
to improve the SNRs. The calibrated blue- and red-arm spectra are 
pieced together directly without scaling or shifting to yield the final spectra. 
Prominent telluric bands, such as the Fraunhofer A and B bands, 
are not removed from the calibrated spectra.  

The scatter of SRCs derived from individual  
standard stars in a given spectrograph indicates that 
the final, average SRCs have achieved an accuracy of a few percent 
for both the blue- and red-arm of the spectrograph.  
Stellar colours deduced by convolving the flux-calibrated 
spectra with the photometric band transmission curves agree with 
photometric measurements of the XSTPS-GAC, with an average 
difference of 0.02$\pm$0.07 and $-0.04\pm$0.09\,mag for $(g-r)$ and $(g-i)$, respectively. 
The relatively large offset in $(g-i)$ is due to the fact that the 
telluric bands in the LSS-GAC spectra, most notably 
the atmospheric A-band in the wavelength range of photometric $i$-band, 
have not been removed from the calibrated LSS-GAC spectra. 
A direct comparison of spectra obtained at multi-epochs 
of duplicate targets indicates that for a spectral SNR per pixel 
($\sim$\,1.07\,{\AA}) at 4650\,{\AA} higher than 30, 
an accuracy of about 10 per cent in the relative flux calibration 
has been achieved for the wavelength range 4000 -- 9000\,\AA.
The accuracy of the calibration degrades dramatically near the 
edges of the wavelength coverage, especially near the short wavelength 
edge of the blue-arm spectra, where the instrumental sensitivity 
drop to a very low value. 
Comparison with the SDSS spectra of common objects shows that 
the LSS-GAC spectra calibrated with the current algorithm exhibit 
more realistic SEDs than the SDSS spectra do for stars from high 
extinction regions.  

The shapes of SRCs are found to show significant temporal variations 
from night to night, even on a single night. 
The variations can reach 30 per cent or more for the whole wavelength range. 
The presence of such large variations in the shapes of SRCs suggests 
that deriving SRCs plate by plate is essential 
for accurate flux calibration of the LAMOST spectra. 
The causes of variations in the shape of SRCs are not very clear. 
Possibilities include changes in fiber response as a result 
of fiber repositioning or erratic variations in the atmospheric reddening 
and extinction. More studies are needed to clarify the situation.

\vspace{7mm} \noindent {\bf Acknowledgments}{ MSX thanks J.-J. Chen for useful discussion. 
This work is supported by NationalKey Basic Research Program of China 2014CB845700. 
Guoshoujing Telescope (the Large Sky Area Multi-Object Fiber 
Spectroscopic Telescope LAMOST) is a National Major Scientific
Project built by the Chinese Academy of Sciences. Funding for
the project has been provided by the National Development and
Reform Commission. LAMOST is operated and managed by the National
Astronomical Observatories, Chinese Academy of Sciences. 
We have used the SDSS spectra and spectra of flux standard stars 
retrieved from the ESO website.}

\end{document}